\documentclass[10pt,aps,twocolumn,showpacs,pra]{revtex4}
\usepackage{amssymb,amsfonts,amsmath,wrapfig,mathrsfs,mathbbol}
\usepackage{graphicx}

\begin{document}

\title{Two-center Interferences in Photoionization of Dissociating H$_2^+$ Molecule}

\author{A. Pic\'{o}n$^1$, A. Bahabad$^{1,2}$, H.C. Kapteyn$^1$, M.M. Murnane$^1$, and A. Becker$^1$}

\affiliation{
$^1$ JILA and Department of Physics, University of Colorado at Boulder, Boulder, Colorado 80309-0440, USA\\
$^2$ Department of Physical Electronics, School of Electrical Engineering, Tel-Aviv University, Tel-Aviv, 69978, Israel}
\date{\today}

\begin{abstract}
We analyze two-center interference effects in the yields of ionization of a dissociating hydrogen molecular 
ion by an ultrashort VUV laser pulse. To this end, we performed numerical simulations of the time-dependent 
Schr\"odinger equation for a H$_2^+$ model ion interacting with two time-delayed laser pulses. The scenario 
considered corresponds to a pump-probe scheme, in which the first 
(pump) pulse excites the molecular ion to the first excited dissociative state and the 
second (probe) pulse ionizes the electron as the ion dissociates. The results of our numerical 
simulations for the ionization yield as a function of the time delay between the two pulses 
exhibit characteristic oscillations due to interferences 
between the partial electron waves emerging from the two protons in the dissociating hydrogen molecular ion. 
We show that the photon energy of the pump pulse should be in resonance with the $\sigma_g - \sigma_u$ transition and the
pump pulse duration should not exceed 5 fs in order to generate a well confined nuclear wavepacket.
The spreading of the nuclear wavepacket during the dissociation is found to cause a decrease of the amplitudes of the
oscillations as the time delay increases. We develop an analytical model to fit the oscillations and show how 
dynamic information about the nuclear wavepacket, namely velocity, mean internuclear distance and spreading, 
can be retrieved from the oscillations. The predictions of the analytical model are tested well against the results of our 
numerical simulations.
\end{abstract}
\pacs{33.80.Rv, 42.25.Hz, 42.50.Hz}

\maketitle

\section{Introduction}
Double-slit like interferences similar to those observed by Young in his experiment with light \cite{young1804}
occur also in the (photo-)ionization of diatomic molecules. Conceptually, the partial electron waves
ejected from the two atomic centers of the molecule take the role of the coherent light waves
emerging from the two holes in Young's experiment. 
The phenomenon was first discussed by Cohen
and Fano \cite{Interference_Cohen_1966} to explain the observation of
oscillations in the total photionization cross section
of the H$_2$ molecule as a function of the photon energy \cite{samson1965}.
According to their analysis, the interference pattern depends periodically on the ratio of the
internuclear distance to the wavelength of the photoelectron and is superimposed on the quickly
decreasing photoabsorption cross section.
In recent years, Young-type
interferences have been observed experimentally and
studied theoretically for the photoemission of (core) electrons from H$_2^+$ and H$_2$
\cite{fojon04,yudin06,fernandez07,fernandez09a,dellapicca09,fernandez09,hu09}
and heavier molecules \cite{rolles05,semenov06,liu06,cherepkov10}.
Furthermore, interference effects in the ionization of molecules by heavy ion \cite{Evidence_Stolterfoht_2001,misra04}
and electron impact \cite{stia03,kamalou05} have been investigated.
Recently, the idea of Cohen and Fano was
extended to the double photoionization of H$_2$
\cite{walter99,akoury07,kreidi08,horner08,schoffler08},
the photoionization of more complex molecules \cite{rudel02}, and
laser induced multiphoton ionization of molecules \cite{muth00,jaron06}.

In photoionization of a diatomic molecule the double-slit interferences primarily
depend on the momentum
of the photoelectron, $k$, the internuclear distance between the two atoms, $R$, and the angle between
the emission direction of the electron and the molecular axis. 
In past work, photoionization of molecules from the ground electronic state and corresponding interference patterns 
in the angular distributions of the photoelectron in the molecular body fixed frame
was studied. Even the small variations of the internuclear distance in the vibrational states of a 
bound molecule could be observed in the patterns \cite{schoffler08,cherepkov10}.

However, the dependence of the interference patterns on the internuclear distance becomes more apparent in 
the photoionization from dissociating molecules. In a recent set of experiments 
 negative iodine molecules I$_2^-$ were excited by a pump laser pulse to a dissociative state, in which 
the molecule fragments into a negative ion I$^-$ and a neutral atom I \cite{davis03,mabbs05,mabbs05a}. 
During the dissociation the molecule was ionized by a second pump laser pulse and the photoelectron spectra 
were measured. The anisotropy parameter obtained from the photoelectron angular distributions
was found to oscillate as a function of the pump-probe time delay, in
other words as a function of the increasing internuclear distance in the dissociating molecule. As in the
previous examples, the oscillation arises due to the interference between the partial electron waves
emerging from the two fragments.

Very recently, the probe step in such kind of pump-probe experiments, 
namely the photoionization of the dissociating molecule, has been
studied theoretically  using first-order perturbation theory
for the prototype example of a H$_2^+$ molecular ion, where the positions of the nuclei were fixed at different 
internuclear separations \cite{chelkowski10}.
Assuming a nuclear wave packet of Gaussian shape with a certain but fixed width $\Delta R$ centered
at an internuclear distance $R_0$,
the angular and momentum distribution of the photoelectron emitted
from the first excited electronic state ($\sigma_u$-orbital) of H$_2^+$ was calculated.
The two-slit interferences were explored for different
orientations of the molecular axis with respect to the polarization direction of the probe laser.

\begin{figure}
\centering\includegraphics[width=7cm]{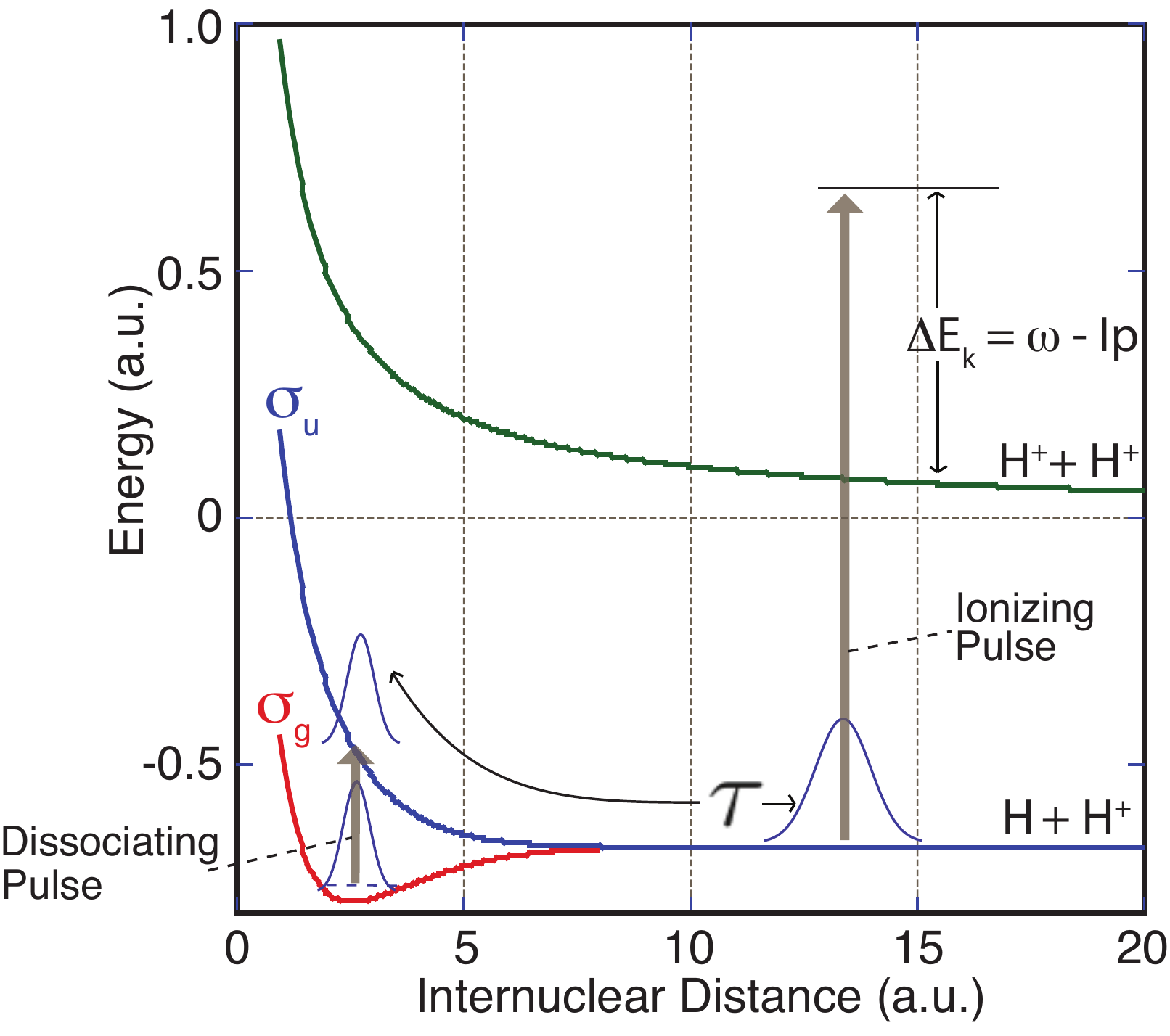}
\caption{
(Color online) 
Scheme of the pump-probe scenario, in which a first (pump) pulse initiates the dissociation by excitation of a wavepacket to the $\sigma_u$-state of H$_2^+$. With a time delay $\tau$ a second (probe) pulse ionizes the dissociating molecular ion. The green curve denotes the energy of two protons as a function of internuclear distance.}
\label{fig:Schematic_Scenario}
\end{figure}

In the present work we complement this recent theoretical study by results of numerical simulations
in which we investigate the complete pump-probe scenario according to the following scheme
(see Fig.\ 1 for illustration).
We consider a
H$_2^+$ molecular ion initially prepared in its electronic ($\sigma_g$) and vibrational ($\nu =0$)
ground state. The interaction with an ultrashort pump pulse resonantly excites part of the wave packet into the 
repulsive $\sigma_u$-state, thus initiating 
the dissociation of the molecule. The ungerade state corresponds to an antisymmetric 
electronic wave function consisting of a coherent superposition of atomic orbitals localized 
at the two dissociating protons, which establishes a double-slit scenario.
During the dissociation the protons first accelerate in
the steep part of the molecular ion potential and then propagate with almost constant velocity. 
Furthermore, the width of the nuclear wave packet 
increases as a function of internuclear distance due to dispersion. This requires to generate a well-confined 
nuclear wave packet in the excitation (pump) step in order to resolve the interferences in the ionization yields, 
as we will show below.
In our simulations, we then modeled the interaction of the dissociating molecular ion 
with a second ultrashort probe laser pulse to
determine the ionization probability as a function of the time delay between the 
dissociating and ionizing pulses. As we will show below, our results exhibit the expected Young-type 
interferences. We finally develop an analytical model in order to use such observations as an imaging tool 
for the velocity, internuclear distance and the spread of the nuclear wave packet. 

The paper is organized as follows. In section \ref{sec:Numerical_Model} we introduce 
the H$_2^+$ model ion used to perform the numerical simulations. Then, in section 
\ref{sec:Dissociation} we study the dissociation of the H$_2^+$ model ion by the pump pulse 
and discuss the analogy of the probability density distribution of the dissociating system to 
a two-slit scenario for different parameters of the pump pulse. 
Next, results for the ionization yields will be presented in section \ref{sec:Photoionization}, 
where the double-slit interference patterns will be identified and analyzed in view of 
imaging the characteristic features of the nuclear wave packet. Finally, we 
conclude with a short summary.


\section{H$_2^+$ model for numerical simulations} \label{sec:Numerical_Model}

For our analysis of the ionization of the dissociating hydrogen molecular ion following the pump-probe
scheme, illustrated in Fig. 1, we make use of a model for H$_2^+$ in which the electronic and nuclear
motion are restricted to one dimension. The model accounts for the interaction of the electron with
the external fields as well as for non-Born-Oppenheimer effects. In the model, the polarization direction of the
two linearly polarized pump and probe laser pulses coincide with the internuclear axis of the molecular
ion as well as with the restricted electronic motion. The field-free
Hamiltonian of this model system is given by (we use Hartree atomic units, $e=m=\hbar=1$)
\cite{hiskes61,kulander96}:
\begin{eqnarray} \nonumber
\hat{H}_{0} &=&
\frac{\hat{P}^{2}}{2\mu_{R}}
+ \frac{\hat{p}^{2}}{2\mu_z}
+\frac{1}{\sqrt{R^{2} + \alpha_{p}}}
\\ 	
\label{Free_H}
&&
- \frac{1}{\sqrt{(z-R/2)^{2} + \alpha_{e}}} - \frac{1}{\sqrt{(z+R/2)^{2} + \alpha_{e}}}
\end{eqnarray}
where $\hat{P}= -i\partial/\partial R$ and $R$, and $\hat{p}= -i\partial/\partial z$ and $z$ are the linear momentum operators and the positions of the relative nuclear (protons) and electronic coordinate, respectively. $\mu_R = M/2$ and $\mu_z = 2M/(M+1)$ are the reduced masses where $M=1836$ a.u. is the mass of the proton. We have chosen soft-core parameters $\alpha_e=1$ and $\alpha_p=0.03$ \cite{kulander96}. The interaction of the electrons is taken into account in dipole approximation and velocity gauge by:
\begin{equation}
\hat{V}_{I} (t) = - (A_d(t)+A_i(t-\tau)) \hat{p}
\end{equation}
with
\begin{eqnarray} \label{Vector_Potential}
A_{l}(t)= \left( 1 + \frac{1}{1+2M} \right)  A_{0,l} \sin(\omega_{l} t) \sin^{2}(\pi t/N_{l}T_{l})
\end{eqnarray}
where $A_{0,l}$ is amplitude of the vector potential, $\omega_{l}$ is the frequency of the carrier wave, $N_{l}$ and $T_{l}$ are the number of cycles and the period of the carrier wave of either the dissociating ($l=d$) or the ionizing ($l=i$) pulse, respectively, and $\tau$ is the time delay between the two pulses.

The corresponding time-dependent Schr\"odinger equation
\begin{eqnarray} \label{Schrodinger_Eq}
i \frac{\partial \psi (z,R,t)}{\partial t} = \left[ \hat{H}_{0} + \hat{V}_{I}(t) \right] \psi (z,R,t) \; ,
\end{eqnarray}
was solved using the Crank-Nicolson method \cite{crank47}. The sizes of the simulation box were
1 to 37 a.u. and -409.6 a.u. to 409.6 a.u. in $R$ and $z$ direction, respectively, with a grid spacing of
$\Delta R$=0.03 a.u. and $\Delta z$=0.1 a.u.. The time step in the simulations was $\Delta t$=0.02 a.u..
The electronic-vibrational ground state of the model hydrogen molecular ion was obtained using
imaginary time propagation, the ground state energy was -0.77 a.u., the equilibirum distance 2.6 a.u. and the
ionization potential for $R \rightarrow \infty$ about 0.71 a.u..

We have chosen to restrict the electronic motion to one dimension along the polarization axis and the
internuclear axis in our model, since we analyzed the ionization of the dissociating hydrogen molecular ion 
up to rather large internuclear distances (37 a.u.) and used a largely extended grid in $z$-direction ($> 800$ a.u.)
in order to keep the wavefunction on the gird for further analysis. 
In recent studies, higher-dimensional calculations have
been performed, in which the full three-dimensional electronic motion \cite{hu09} or the two-dimensional 
electronic motion in
cylindrical coordinates have been taken into account \cite{chelkowski95,roudnev04,he07,he08}. These
higher-dimensional calculations are usually either performed on smaller grids or make use of high performance
supercomputers. Here, we are interested in the analysis of the double-slit interference effects in the ionization
signal from the dissociating molecular ion as a function of the internuclear distance $R$. These effects
are expected to be present independent of the emission direction of the electron, which justifies a
restriction of the electronic motion to one dimension.


\section{Dissociation dynamics} \label{sec:Dissociation}

\begin{figure}
\centering\includegraphics[width=7cm] {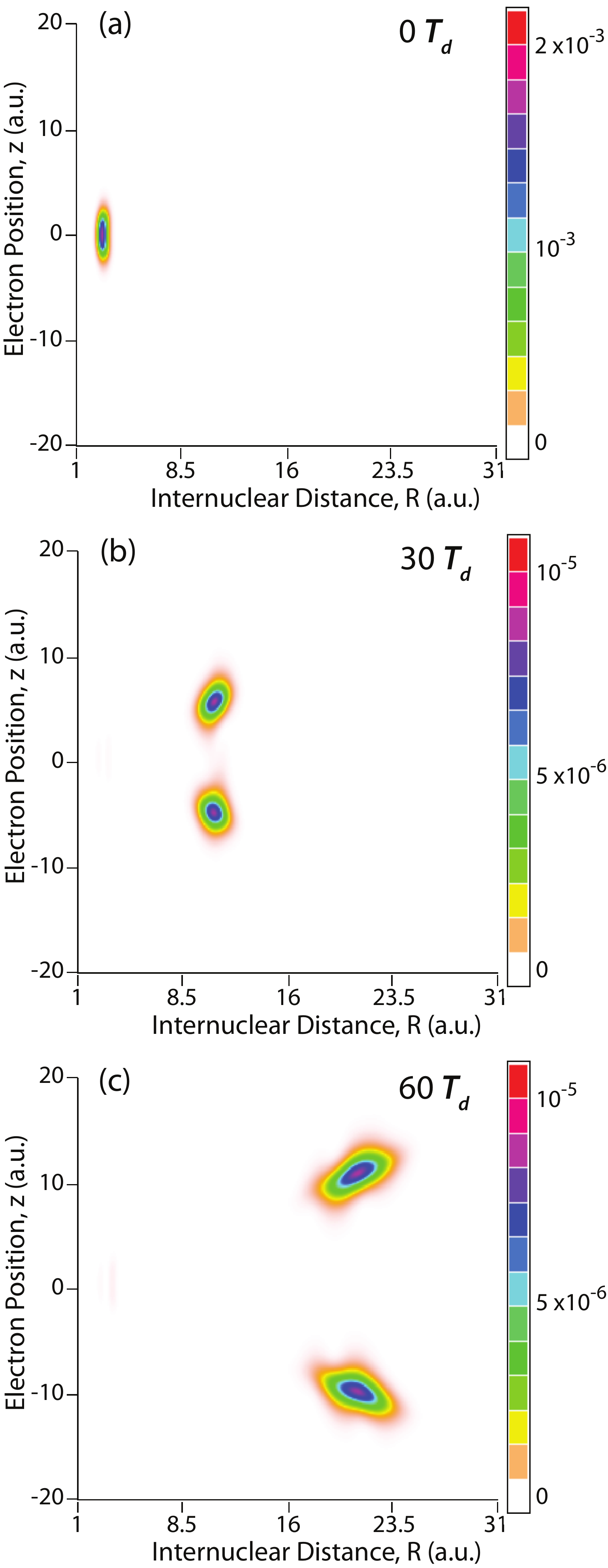}
\caption{(Color online) 
Comparison of the electron probability density in the ground state of the H$_2^+$ model (panel (a)) with 
the temporal evolution of the density in the first excited state of hydrogen molecular ion following an excitation 
of the molecular ion 
by an ultrashort laser pulse with a central wavelength of 117 nm and a full pulse length of 48.66 a.u. (about 1.2 fs). 
Shown are snapshots of the excited state wavepacket density at 
(b) $30 T_{d}$ (11.782 fs) and (c) $60 T_{d}$ (23.564 fs).
}
\label{fig:Dissociation}
\end{figure}

First, we analyze the dissociation dynamics of the hydrogen molecular ion after the interaction with the 
pump laser pulse. To this end, we performed numerical simulations of the model hydrogen molecular ion 
interacting with just one ultrashort laser pulse. We filtered out the electronic-vibrational ground state 
after the interaction and monitored the wavepacket in the excited state as a function of time. 
In Fig. \ref{fig:Dissociation} we compare two snapshots of the temporal evolution of the excited state 
probability density as a function of $z$ and $R$ (panels (b) and (c)) with the corresponding distribution 
in the ground state (panel (a)). The calculations were performed for a dissociating pump pulse with $\omega_{d}=0.38$ a.u. (period $T_{d}=16.22$ a.u.), 
corresponding to a central wavelength of 117 nm, which is resonant with
the $\sigma_g - \sigma_u$ transition in our model. The full pulse contained three cycles ($N_l =3$), i.e.\
a full pulse length of 48.66 a.u., about 1.2 fs. 
The amplitude of the vector potential was $A_{0,d}=0.04$ a.u., 
corresponding to a peak intensity of $8.4\times10^{12}$ $\rm W/cm^{2}$. 
At the end of the pulse, the total excitation probability, i.e. the probability after 
projecting out the $\sigma_{g}$ ground state, was 2.26\%.

Fig. \ref{fig:Dissociation} shows, as expected, that the excited state probability density 
propagates to larger internuclear distances after the interaction with the pump pulse, 
which indicates a dissociation of the molecular ion. One sees that, at a given internuclear distance, 
the density is almost equally (de-)located at both protons and, hence, it forms the anticipated set-up, similar to a double slit. We note that the initially very well confined nuclear wave packet with a narrow distribution in $R$ does spread significantly due to dispersion as the time proceeds.

\begin{figure}
\centering\includegraphics[width=7cm] {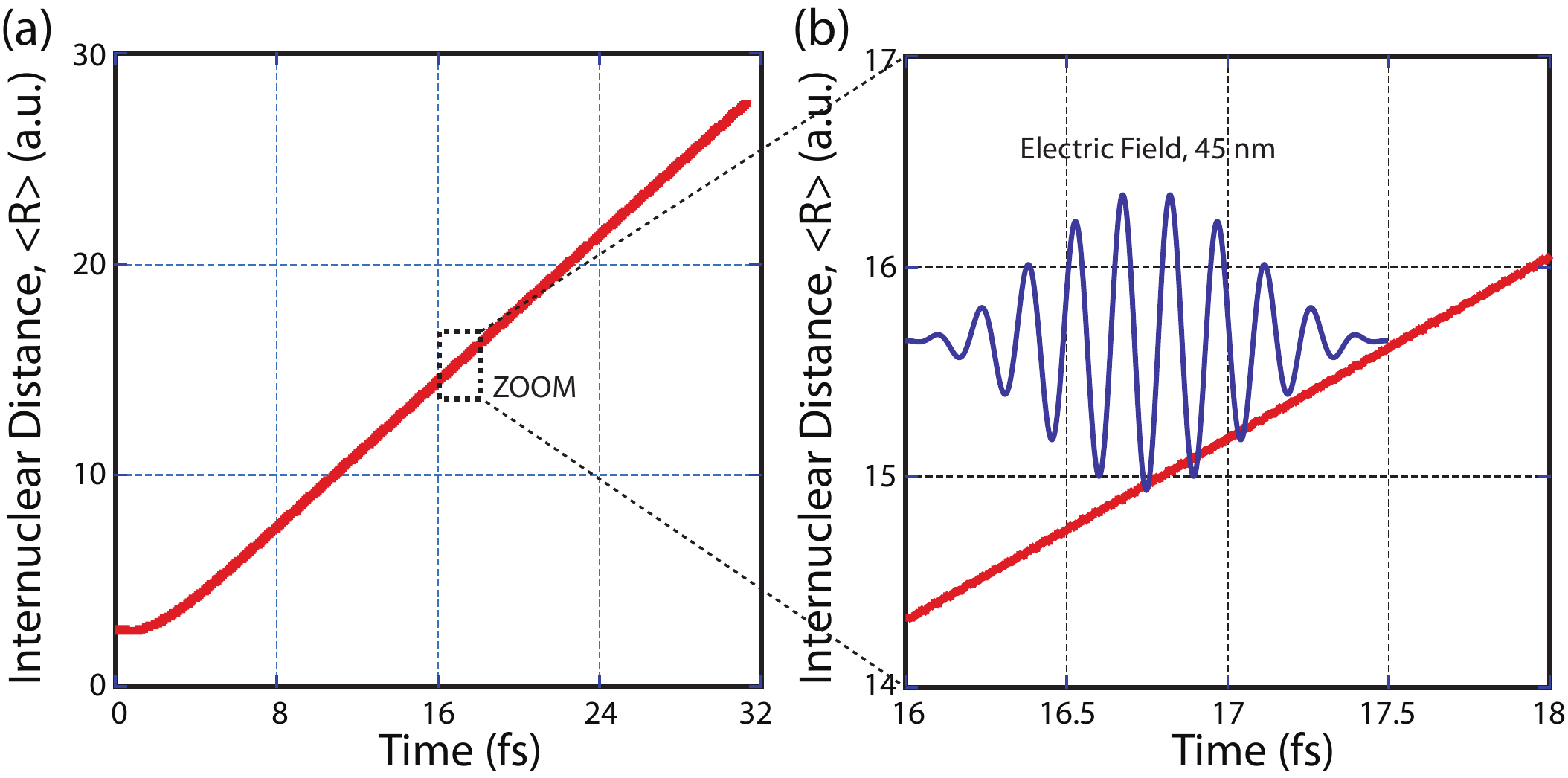}
\caption{(Color online) 
(a) Expectation value of the internuclear distance as a function of time. 
(b) Enlarged view of (a) for a time interval of 2 fs (16 fs to 18 fs).
For the sake of comparison, we have also plotted the electric field of a 10-cycle pulse at a wavelength of 45 nm, which was used as a probe pulse below.}
\label{fig:Mean_Value_R}
\end{figure}

Next, we present in Fig. \ref{fig:Mean_Value_R} the result from the numerical simulations for the 
expectation value of the internuclear distance $\langle R\rangle$ as a function
of time after the interaction with the pump pulse. Please note that the expectation value was calculated
after the remaining population in the ground state was removed. It is clearly seen that the wave packet
gets accelerated over the first few femtoseconds due to the steep gradient of the repulsive potential
energy curve for the $\sigma_u$-state at short internuclear distances. However, we observe that about
4-5 fs after the end of the interaction with the pump pulse the velocity of the wave packet is almost
constant and $\langle R\rangle$ increases linearly with time. From the expanded view in Fig. \ref{fig:Mean_Value_R}(b) we notice
that the expectation value $\langle R\rangle$ changes by $0.86$ a.u. ($\pm0.02$ a.u.) within 1 fs in the linear regime, i.e.\ the 
nuclear wave packet propagates with a velocity of $0.0104$ a.u. ($\pm0.0003$ a.u.).

\begin{figure}
\centering\includegraphics[width=9cm] {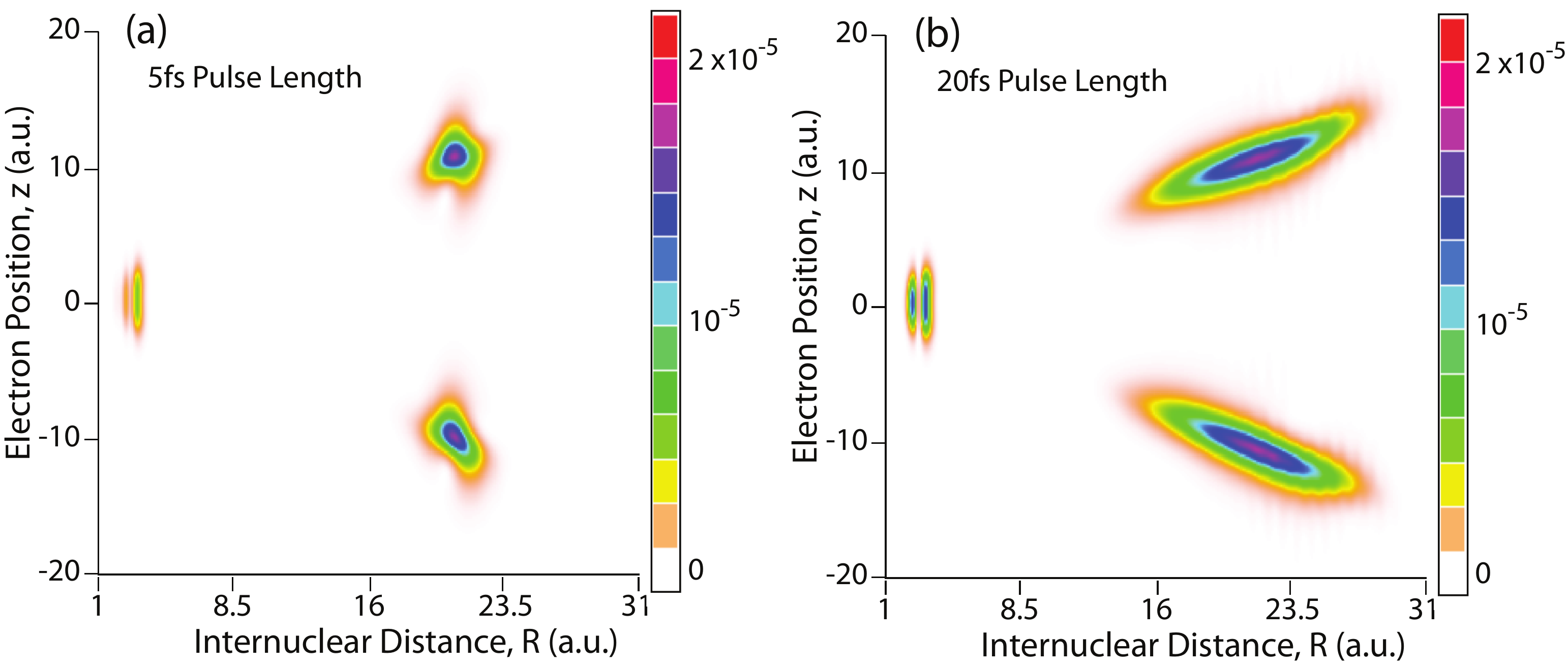}
\caption{(Color online) 
Electron probability density (after removal of the electronic-vibrational ground state) following 
the interaction with pump laser 
pulses at 117 nm having a full pulse length of (a) 5 fs at $t=61 T_{d}$ (23.912 fs) and (b) 20 fs at $t=80 T_{d}$ (31.360 fs).
}
\label{fig:Dissociation_2}
\end{figure}

We also studied the dissociation dynamics for pump pulses with longer pulse lengths as well as with 
different frequencies in order explore the parameter regime of the pump pulse for creating a well-confined two-center scenario in the dissociating molecular ion. 
In Fig. \ref{fig:Dissociation_2} we present the excited state probability density generated 
by different pump pulses of 5 and 20 fs, which are resonant with the $\sigma_g - \sigma_u$ transition. 
While the wavepacket generated with the 5 fs pulse is still well confined at large internuclear distance, we observe 
that the spread of the wavepacket is significantly larger after excitation with a 20 fs pulse. 
This is due to the fact that for longer pulses, the (effective) interaction time between the molecular ion and 
the pump pulse increases. Consequently, the width of the initial wavepacket created in the excitation step 
increases, which leads to a wider wavepacket at larger internuclear distances. 
Please note that both these pulses create a significant amount of vibrational excitation in the 
bound molecular ion (cf. density at small internuclear distances in Fig. \ref{fig:Dissociation_2}).

\begin{figure}
\centering\includegraphics[width=9cm] {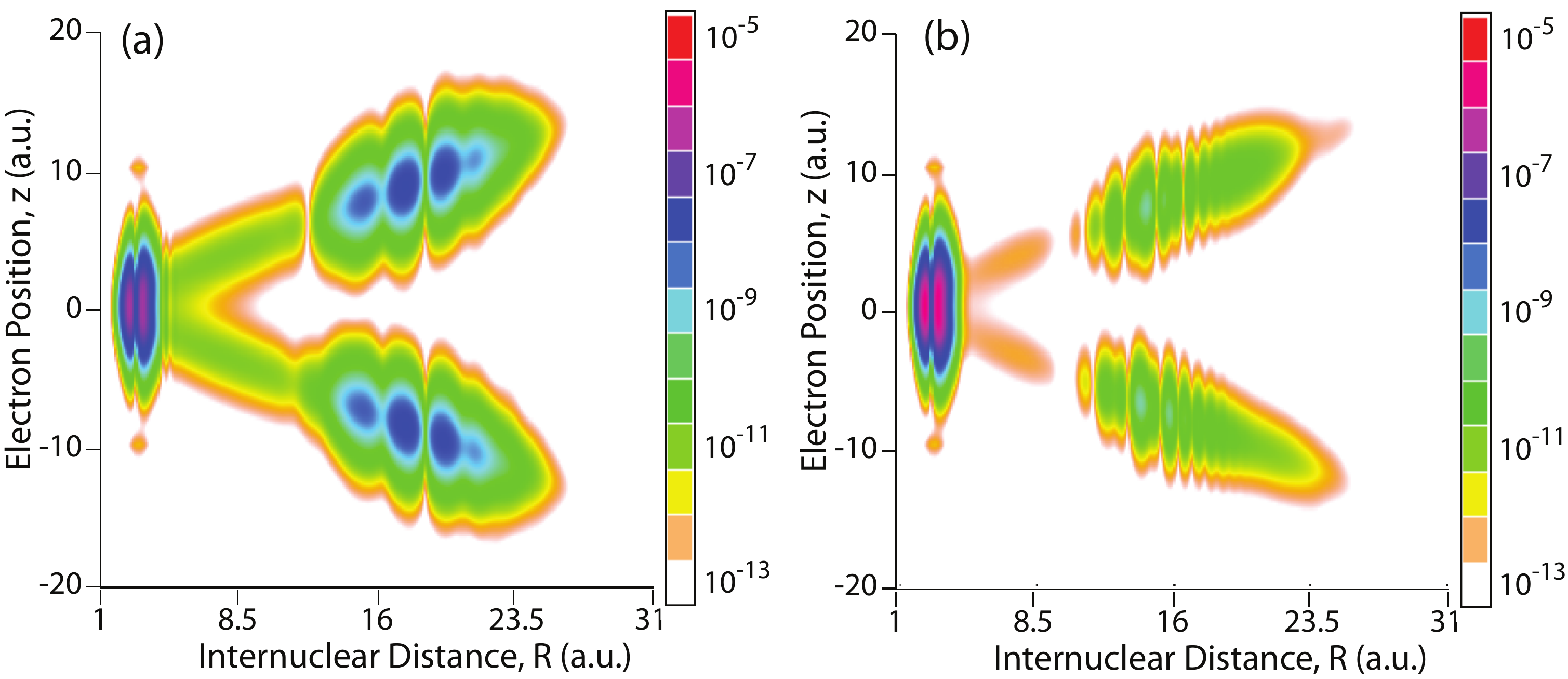}
\caption{(Color online) 
Electron probability density (after removal of the electronic-vibrational ground state) following 
the interaction with pump laser pulses at 800 nm with a peak intensity around 10$^{13}$ W/cm$^{2}$ and
having a full pulse length of (a) 1 cycle (i.e.\ 2.7 fs) at 23.564 fs and (b) 3 cycles (i.e.\ 8.1 fs) at 25.872 fs.
}
\label{fig:Dissociation_3}
\end{figure}

For the sake of comparison, we show in Fig. \ref{fig:Dissociation_3} 
the corresponding excited state probability densities generated by the 
interaction with a non-resonant pump laser 
pulse operating at 800 nm and total pulse lengths of (a) 1 cycle (2.7 fs) and (b) 3 cycles (8.1 fs). 
At this wavelength it is well-known that the dissociation of H$_2^+$ occurs via 
other mechanisms such as above threshold dissociation \cite{suzor90}, 
bond softening \cite{bandrauk81,bucksbaum90}, or bond hardening \cite{frasinski99}. 
Instead of one (narrow or wide) wavepacket, as in the case of a resonant one-photon 
excitation with the VUV pulses, we observe a train of nuclear wave packets for both pulse lengths. 
Also, in this case the vibrational states are much more excited than in case of an excitation by the VUV pulses.

We may summarize our results so far in view of our goal to observe double-slit interference effects 
from the dissociating hydrogen molecular ion. For this goal, we require a single well-confined 
nuclear wavepacket, which corresponds to a narrow distribution of the internuclear distances 
(which shows certain similarities with a double-slit having a well-defined separation of the slits). From the results presented above, 
we conclude that the system should be prepared in the vibrational ground state, the pump pulse should be in resonance with the $\sigma_g - \sigma_u$ transition 
(in order to avoid trains of wavepackets as in the case of dissociation driven by the 800 nm pulses) 
and as short as possible (in order to confine the initial wavepacket as much as possible). 
Furthermore, we infer from our result for the velocity of the nuclear wavepacket (cf. Fig. \ref{fig:Mean_Value_R}(b)) 
that the ionizing probe pulse should be less than about 2 fs. Otherwise the wavepacket sweeps over a rather 
large region of internuclear distances during the interaction with the probe pulse which will likely 
smear out any interference patterns in the ionization signal, as long as the internuclear distance at 
the moment of ionization is not detected in the experiment.


\section{Photoionization of dissociating H$_2^+$ molecule} \label{sec:Photoionization}

In this section we present the results of numerical simulations in which we took the complete pump-probe scheme into account (Fig. \ref{fig:Schematic_Scenario}). First, we focus on the identification of interference patterns in the yields of ionization of the dissociating hydrogen molecular ion by a time-delayed probe pulse and then we analyze the results in view of an imaging of the nuclear wave packet. For our analysis we used an ultrashort dissociating pump pulse, resonant with the $\sigma_g - \sigma_u$ transition ($\omega_d$=0.38 a.u., $\lambda$=117 nm) and a full pulse length of 48.66 a.u. (1.2 fs, or 3-cycles) and a peak intensity of $8.4\times 10^{12}$ W/cm$^2$. The corresponding dissociation dynamics of the model H$_2^+$  is represented by the results shown in Figs. \ref{fig:Dissociation} and \ref{fig:Mean_Value_R}. In each simulation, we project out the ground state of the H$_2^+$ model ion after the interaction with the pump pulse,
assuming that any ionization signal from the bound molecular ion can be well separated in the experiment.

\subsection{Double-slit interferences}

We performed a series of numerical simulations in which we considered the application of a second photo-ionizing
probe pulse as a function of time delay $\tau$. After the end of the probe pulse 
we calculated the ionization yield via the integral over the outgoing 
parts of the electron wave packet as (please note, that we projected out the ground state of the H$_2^+$ model after the pump step):
\begin{eqnarray}
P^+ = \lim_{t \rightarrow \infty} \int\limits_{0}^{\infty} dR \int\limits_{|z|=10+\frac{R}{2}}^{\infty}dz |\Psi(z,R,t)|^2\, . 
\end{eqnarray}
We continued the propagation of the wavefunction until the ionization signal determined on the grid did not 
change as a function of time.

According to previous analysis the photoionization yield from a diatomic molecule is expected to be modulated as
\cite{Interference_Cohen_1966,liu06}:
\begin{eqnarray} \label{Oscillation1}
P^{+} ({\bf k},{\bf R}) \propto 1 + \cos({\bf k}\cdot {\bf R} + \Phi({\bf k},{\bf R})) \; ,
\end{eqnarray}
with $\Phi({\bf k},{\bf R}) = \phi_{\rm orb} + \phi_{\rm scat}({\bf k},{\bf R})$, where $\phi_{\rm orb}$ accounts for the phase of the symmetry of the orbital from which the electron is emitted ($\phi_{\rm orb} = \pi/2$ for the present case, as the electron is in the $\sigma_u$-orbital), and $\phi_{\rm scat} ({\bf k},{\bf R})$ represents a phase acquired due to the scattering of the electron by the neighboring proton \cite{liu06}. Since the symmetry of the electronic molecular wavefunction does not change during the dissociation, $\phi_{\rm orb}$ does not depend on the internuclear distance and the photoelectron momentum. In contrast, one may expect that in general the acquired scattering phase depends on both these parameters.
Since in our calculations we restricted the electron motion to one dimension along the internuclear axis of the molecular ion, Eq.\ (\ref{Oscillation1}) reduces to:
\begin{eqnarray} \label{Oscillation}
P^{+} (k,R) \propto 1 + \cos(kR + \Phi(k,R)) \; .
\end{eqnarray}

We considered two different ionizing pulses. The first one had a frequency of 
$\omega_{i}=2.07$ a.u. (56.33 eV per photon), corresponding to a wavelength of 22 nm and a period of $T_{i}=3.03$ a.u. (73 as), $N_{i}=10$ 
(i.e., full pulse length of 730 as) and $A_{i}=0.03$ a.u. (i.e., peak intensity of $1.35\times10^{14}$ $\rm W/cm^{2}$). 
The parameters of the second pulse were 
$\omega_{i}=1.01$ a.u. (27.54 eV per photon), corresponding to a wavelength of 45 nm and a period of $T_{i}=6.20$ a.u. (150 as). 
The number of cycles was $N_{i}=10$ (i.e., full pulse length of 1.5 fs) and the amplitude of the vector potential was $A_{i}=0.06$ a.u. 
(i.e., peak intensity of $1.3\times10^{14}$ $\rm W/cm^{2}$).

\begin{figure}
\centering\includegraphics[width=5cm] {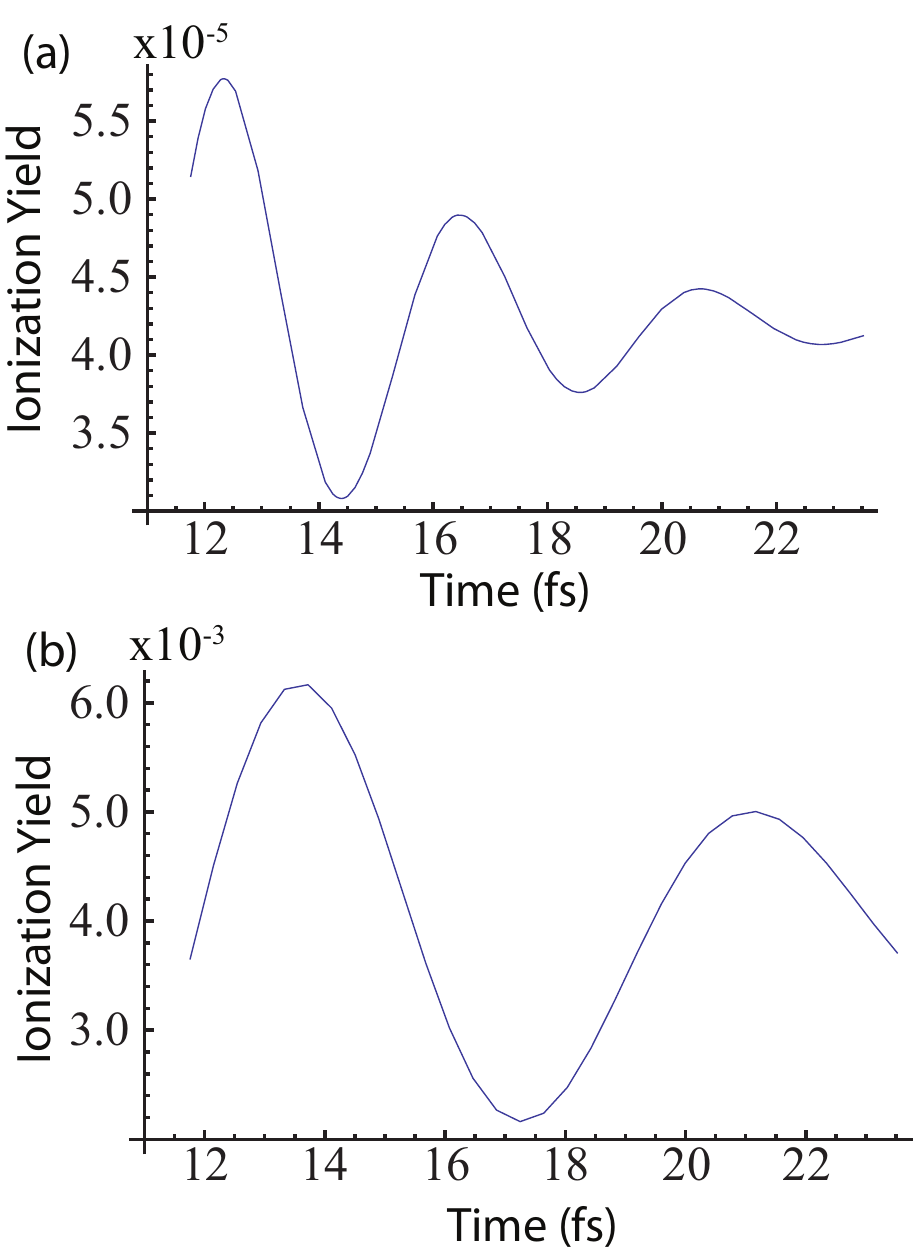}
\caption{(Color online) 
Results of numerical simulations for the ionization yield using two different 
wavelengths for the ionizing pulse: (a) 22 nm and (b) 45 nm.}
\label{fig:Atto_Atto_Interferences}
\end{figure}

In Fig. \ref{fig:Atto_Atto_Interferences} we show the numerical results for the 
ionization yield as a function of the pump-probe time delay for the two ionizing pulses.
We clearly observe oscillations of the ionization yields due to the interferences of the partial electron wave packets. The amplitudes of the oscillations decrease, which we attribute to the spreading
of the nuclear wavepacket (for further analysis, see section IV.B.). Over the range of time delays studied, the frequency of the oscillations remains practically constant 
($0.020 \pm 0.001$ a.u. for ionization at 45 nm, and $0.038 \pm 0.005$ a.u. at 22 nm). Since the ionization potential of the dissociating hydrogen molecule in the present range of 
internuclear distance does vary very little and, hence, the photoelectron momentum remains almost constant, we conclude that the phase $\Phi$ and, hence, the scattering phase 
$\phi_{\rm scat}$ is independent of the internuclear distance $R$ (or, of the time delay $\tau$). This result makes sense, since in the present model we restricted the electron 
motion to one dimension along the internuclear distance and at large internuclear distances the Coulomb potentials located at the two protons are well separated and vary just 
slightly with an increase of $R$. Thus, the scattering process of the electron at the neighboring Coulomb potential is (almost) independent of the value of $R$ and, hence, 
$\Phi(k,R) = \Phi(k)$ for large $R$ in our model.

Before we proceed, we may mention that we have also considered probing the dissociating molecular ion using
near-infrared laser pulses at 800 nm. Due to the non-perturbative coupling between 
the electron and field at this wavelength, the 
ionization of the electron proceeds via tunneling or multiphoton above-threshold 
ionization. Consequently, the electron is not emitted at a given energy (within a certain width) but with a 
range of electron energies that is rather broad. In order to observe any trace of the double-slit interference 
as a function of the internuclear distance, it is therefore necessary to restrict the electron energy in an 
experiment as well as in the numerical simulations. In other words, instead of total ionization yields, we have 
to obtain partial differential ionization yields 
\begin{equation}
\Delta P^{+}=(dP^+/dk) \Delta k \; .
\end{equation}
Furthermore, the period of a near-infrared field 
is about 2.7 fs, over which the expectation value of the nuclear wavepacket changes by more than 2 a.u. 
(see, Fig. \ref{fig:Mean_Value_R}).  As 
long as the internuclear distance is not (indirectly) observed in an experiment, any interference effects
will therefore 
smear out for near-infrared probe pulses with a few cycles of duration. Finally, strong-field ionization 
effects, such as rescattering \cite{corkum93} and multiple ionization bursts per half cycle of the field 
\cite{takemoto10}, are other obstacles for the occurence of the double-slit interference effects in 
the partial differential ionization yields.

\begin{figure}
\centering\includegraphics[width=5cm] {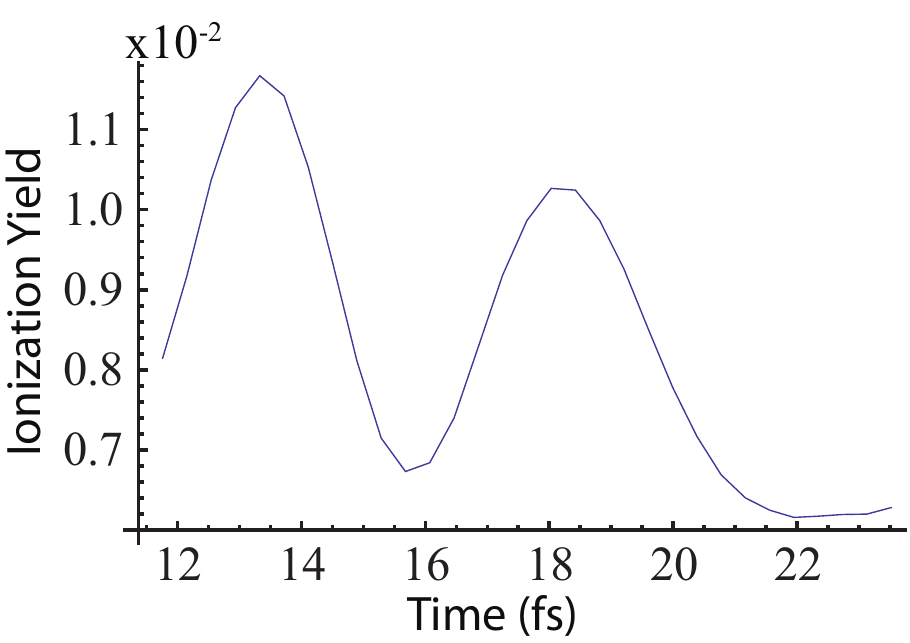}
\caption{(Color online) 
Results of numerical simulations for the 
partial differentail ionization yield of H$_2^+$ interaction with a 1-cycle pulse at 800 nm. 
The photoelectron linear momentum was selected as 
$k_{0}=-1.36$ a.u. and a detector resolution of $\Delta k$= 0.1 a.u. was considered.}
\label{fig:800nm}
\end{figure}

Assuming an ideal scenario, in which the dissociating hydrogen molecular ion is ionized by an 800 nm 
laser pulse with just one cycle (total pulse length of 2.7 fs), we have performed numerical simulations for a 
pulse with a peak of the vector potential of $A_{0,i}=3$ a.u. (about $10^{15}$ $\rm W/cm^{2}$ peak intensity). 
As shown in Fig. \ref{fig:800nm}, we observe a characteristic oscillation in the partial differential 
ionization yield at a specific value of the 
photoelectron momentum of  $k_{0}=-1.36$ a.u.. The negative sign indicates that the electron is emitted 
in direction opposite to the laser polarization direction. We note that the yield for electrons emitted 
with the same absolute momentum value in the 
laser polarization direction, as well as the yields at different electron momenta, 
do not show any characteristic 
oscillatory behaviour. Thus, despite the fact that the oscillation in the specific yield in Fig. \ref{fig:800nm} appears to resemble the interference patterns, we expect that in general interference effects cannot be observed using a intense probe laser pulse at near-infrared wavelengths.

\subsection{Imaging of nuclear wavepacket dynamics}

Now, we will analyze which information about the dissociating nuclear wave packet (velocity, spreading, etc.) can be retrieved from the double-slit interference patterns. For large internuclear distances we can well approximate 
\begin{equation}
k R(\tau) + \Phi \approx 2 k v \tau + k R_{c} + \Phi
\end{equation}
where $v$ is the velocity of the dissociating protons and $k R_{c}$ is a phase that depends on the initial 
internuclear distance and the initial acceleration due to the repulsive potential at small internuclear distances. 
Thus, the frequencies of the oscillations in Fig. \ref{fig:Atto_Atto_Interferences} are 
approximately equal to $2 k v$ and we can retrieve the velocity (or momentum) of the protons.
From the results of our simulations for the 45 nm ionizing pulse, 
we read a frequency of $0.020$ a.u. ($\pm 0.001$ a.u.), while the photoelectron 
momentum is $k=0.83$ a.u. ($\pm0.02$ a.u.), obtained from the Fourier transform of the ionized photoelectron wave functions in the 
simulations. 
This gives rise to a velocity of $v=0.012$ a.u. ($\pm 0.001$ a.u.). For the 22 nm ionizing pulse we obtain a frequency of
$0.038$ a.u. ($\pm0.005$ a.u.) and a momentum of $k=1.64$ a.u. ($\pm0.02$ a.u.), resulting in a velocity of $v=0.011$ a.u. ($\pm 0.001$ a.u.).
Both results are in good agreement with the velocity of $v=0.0104$ a.u. ($\pm 0.0003$ a.u.) obtained from the results for the expectation value of the 
internuclear distance $\langle R\rangle$, shown in Fig. \ref{fig:Dissociation}(b). 

As outlined before, the decay of the oscillations in Fig. \ref{fig:Atto_Atto_Interferences} depends on both the spread 
of the nuclear wavepacket in the internuclear distance, which increases with  $\tau$, and the width of 
photoelectron momentum distributions, which remains constant with $\tau$ in the present study. 
To estimate the decay we revise the formula for the double-slit interferences, given in Eq. (\ref{Oscillation}), 
to account for the spreading as:
\begin{eqnarray} 
\lefteqn{
P^+(k_{0},R_{0}) \propto
\int_{-\infty}^{\infty} \int_{-\infty}^{\infty} dk\, dR}
\label{Mk_Simplest_Model}
\\
\nonumber
&&
\left[ 1 + \cos(kR + \Phi)\right]
G(R-R_{0},\Delta R) \, G(k-k_{0},\Delta k)
\end{eqnarray}
where $G(a,b)$ is a Gaussian profile defined as
\begin{eqnarray} \label{Gaussian_Profile}
G(a,b) = \frac{1}{\sqrt{2\pi} \, b}\; e^{-\frac{a^{2}}{2b^{2}}}\; ,
\end{eqnarray}
with $R_{0}\equiv\langle R\rangle$ being the mean value of the internuclear distance 
(see Fig. \ref{fig:Dissociation}(b) for its dependence on $\tau$), 
$\Delta R$ is the width of the nuclear wavepacket (which depends on $\tau$ as well), 
$k_{0}$ and $\Delta k$ are the center and the width of the photoelectron momentum distribution, respectively 
(in an actual experiment, $\Delta k$ may also be related to the resolution of the detector). 
Assuming that $\Phi$ does not depend on the internuclear distance and the central photoelectron momentum is 
constant, we perform the integrations in Eq. (\ref{Mk_Simplest_Model}) 
to get:
\begin{eqnarray} \nonumber
\lefteqn{P^+(k_{0},R_{0})  \propto} 
\\
&&
\left[ 1 \pm \frac{1}{\Delta}\cos\left(\frac{k_{0}R_{0} }{\Delta^{2}}+ \Phi\right) \, e^{-\frac{R_{0}^{2}\Delta k^{2} + k_{0}^{2}\Delta R^{2} }{2 \Delta^{2}}} \right] \; , \label{Mk_Simplest_Model_2}
\end{eqnarray}
where  $\Delta = \sqrt{ 1 + \Delta k^{2} \Delta R^{2} }$. 
$k_{0}$ and $\Delta k$ are parameters, which can be retrieved in an actual experiment (here, we take 
these parameters from the results of our numerical simulations). Thus, we are left with two variables 
in Eq. (\ref{Mk_Simplest_Model_2}), namely the internuclear distance $R_{0}$ and the spreading of 
the nuclear wavepacket $\Delta R$. As explained before, at large internuclear distances the Coulomb repulsion 
between the two protons becomes negligible and the protons propagate with constant momentum. In this regime we 
get 
\begin{equation}
R_{0}(\tau) \approx 2 v \tau + R_{c}
\label{fit1}
\end{equation} 
and 
\begin{equation}
\Delta R (\tau) \approx \sqrt{1+4\tau^2\Delta k_{p}^{4}/m_{p}^2} / (2\Delta k_{p}) + \Delta R_{c}, 
\label{fit2}
\end{equation}
where $\Delta k_{p}$ is the spreading of the relative momentum of the protons and
$m_{p}=918$ a.u. is the reduced mass of the two protons. $\Delta R_{c}$ and $R_{c}$ are two terms which account for 
the initial nuclear spread, the initial internuclear distance and the initial acceleration of the protons, 
where the linear temporal
approximation fails due to the strong repulsive Coulomb potential. As shown before, $v$ can be retrieved with 
high accuracy from the interference pattern. Assuming that $\Delta k_p$ can be either observed in the experiment or 
approximately retrieved from the width of the photoelectron momentum distribution, we need to fit the two parameters 
$R_c$ and $\Delta R_c$ to obtain (the mean value of) the internuclear distance and the width of 
the nuclear wave packet at the moment of ionization from the decay of the oscillations.

\begin{figure}
\centering\includegraphics[width=6cm] {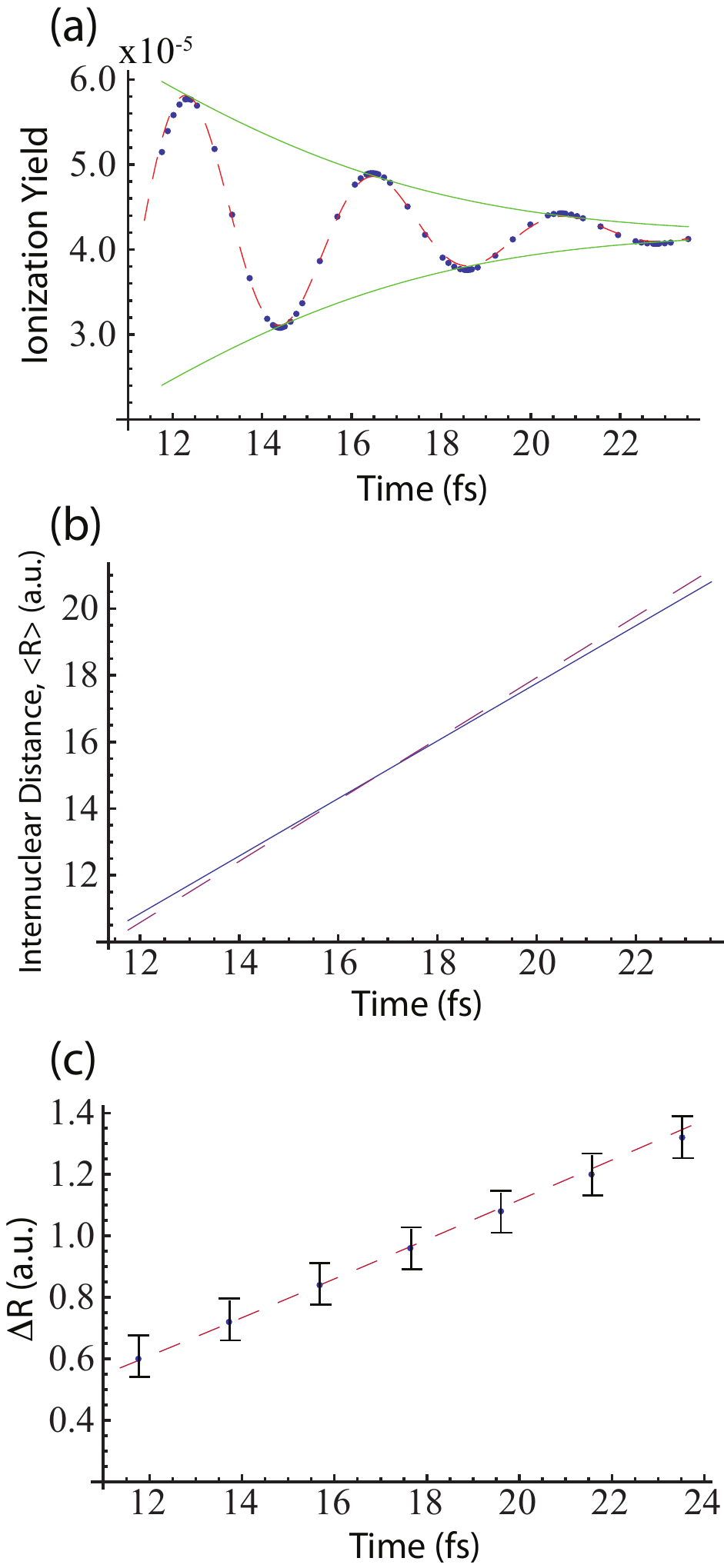}
\caption{\small (Color online) 
(a) Ionization yields: Comparison between the results of numerical simulations (blue points) and the 
predictions of the analytical model, Eq. (\ref{Mk_Simplest_Model_2}), with (red dashed line) and 
without (green solid line) the $\cos$-term, choosing $v=0.0111$ a.u., $k=1.64$ a.u., $\Delta k=0.084$ a.u., 
$\Delta k_{p}=1.48$ a.u., 
$R_{c}=-0.45$ a.u., $\Delta R_{c}=-0.26$ a.u. and $\Phi=1$ a.u..
(b) Mean value of the internuclear distance: Comparison between the results of numerical simulations
(blue solid line) and the fit formula, Eq. (\ref{fit1}) (red dashed line).
(c) Spreading of the nuclear wavepacket:  Comparison between the results of numerical simulations 
(points with error bars due to the grid resolution) and the fit formula, Eq. (\ref{fit2}) (red dashed line).
}
\label{fig:Fitting_22nm}
\end{figure}

In order to check our analytical model, we performed numerical simulations for the 22 nm ionizing pulse 
with a high resolution of the time delay ($\Delta \tau= 0.04, T_{d}= 15$ as) (cf.\ points in 
Fig. \ref{fig:Fitting_22nm}(a)). 
Choosing $R_{c}=-0.45$ a.u. and $\Delta R_{c}=-0.26$ a.u. we obtain a very good 
fit of the decay of the amplitudes of the oscillations 
(see green solid curves in Fig. \ref{fig:Fitting_22nm}(a)). 
Using these parameters and choosing $\Phi$=1 rad, we were also able to fit the detailed oscillations 
very well (red dashed curve). 
As explained above, the fitting procedure allows us to further retrieve the mean value of the 
internuclear distance and the width of the nuclear wave packets. 
In Figs. \ref{fig:Fitting_22nm}(b) and (c), we compare the results of these fits (red dashed lines) 
with the results retrieved from our numerical simulations (blue solid lines and points). 
The agreement is very satisfactory. Therefore, we can expect to be able to extract dynamical information 
about the nuclear wavepacket from such interference patterns in an experiment.


\section{Summary} \label{sec:Conclusions}

We have studied double-slit like interferences in the photoionization yields 
of a dissociating hydrogen molecular 
ion via numerical simulations of the time-dependent Schr\"odinger equation. 
To this end, we analyzed the different steps in a pump-probe scenario, in which H$_2^+$ is excited 
by a pump pulse from the ground state to the first excited dissociative state and is then probed by 
a second time-delayed pulse, which ionizes the electron from the molecular ion as the ion dissociates. 
The results of our numerical simulations for a H$_2^+$ model ion in its vibrational ground state show that the frequency of the pump pulse
should be in resonance with the $\sigma_g - \sigma_u$ transition and should not be longer than about 5 fs 
in duration (total pulse duration) in order to generate a well confined nuclear wavepacket, which is 
a requirement to observe interference effects in the dissociating molecular ion (as long as the 
internuclear distance of the protons at the moment of ionization is not observed). We have shown 
that the yields of ionization due to the interaction with a VUV probe pulse exhibit the expected double-slit 
interferences as a funtion of time delay between the two pulses. The amplitudes of the oscillation are 
found to decrease with increase of the time delay due to the spreading of the wave packet. 
By developing an analytical 
model we were able to retrieve information about the nuclear wavepacket, namely its velocity, the mean 
value of the internuclear distance and the width of the wavepacket, from the oscillations 
in the ionization yields. We have tested the predictions of the analytical model against the results 
of numerical simulations and found a very satisfactory agreement.

\section*{Acknowledgements}

AP acknowledges financial support from the Spanish Ministry of Science and Innovation through their 
Postdoctoral program. The work was partially supported by NSF.

\end{document}